\documentclass[11pt,twoside]{article}
\usepackage{FUSE2004}
\usepackage{natbib}

\usepackage{epsf}
\usepackage{psfig}
\usepackage{lscape}

\begin{document}
\title{FUSE observations of the HeII Lyman alpha forest towards HS~1700+6416}
\author{D. Reimers, C. Fechner, G. Kriss, M. Shull, R. Baade, W. Moos, A. Songaila, R. Simcoe}
\affil{Universit\"at Hamburg, Hamburg, Germany; STSci, Baltimore, MD, USA; University of Colorado, Boulder, CO, USA; Johns Hopkins University, Baltimore, MD, USA; University of Hawaii, Honolulu, HI, USA; MIT, Cambridge, MA, USA}

\begin{abstract}
We present FUSE observations of the \ion{He}{2} Ly$\alpha$ forest 
in the redshift range $2.3 < z < 2.7$ towards HS~1700+6416. Between 
October 2002 and February 2003, the brightness of the 
QSO increased by a factor 2. Therefore, with an exposure 
time of 203\,ks during orbital night, the quality of the 
resulting spectrum is comparable to the HE~2347-4342 data. This 
second line of sight with a resolved \ion{He}{2} Lyman alpha 
forest reveals a similar variation of several orders of magnitude of the 
column density ratio $\eta = N($\ion{He}{2}$)/N($\ion{H}{1}$)$ and confirms the results of 
previous studies. The well-known metal line spectrum of HS~1700+6416 
permits to examine the influence of metal line absorption on 
the \ion{He}{2} column densities.
\end{abstract}

\section{Introduction}

Resolving the \ion{He}{2} Ly$\alpha$ forest in HE~2347-4342 with FUSE \citep{krissetal01} has led to a detailed picture of the ionization if the IGM. 
The cosmic UV background which governs the \ion{He}{2}/\ion{H}{1} column density ratio was found to fluctuate by at least two orders of magnitude, with the highest values observed in the low density part (voids) of the Ly$\alpha$ forest \citep{shulletal04}.
We present here FUSE observations of HS~1700+6416 ($z_{\mathrm{em}} = 2.73$), a second line of sight of comparable quality.

\section{Observations and continuum estimation}

A summary of the observations is given in Tab. \ref{observations}.
The data reduction of the FUSE spectrum was performed by G. Kriss.
The wavelength range of the \ion{H}{1} Ly$\alpha$ forest is covered by two Keck/HIRES spectra \citep{songaila98,simcoeetal02}.
We coadd these two datasets. 
The Simcoe spectrum reaches down to 3200\,\AA\, so this portion is used as well for the analysis of the metal line systems along the line of sight.
Furthermore, we have STIS data from the HST archive, taken with the Echelle E140M grating.
Because of their low signal-to-noise ratio only the best part of the spectrum is used for consistency checks constraining the models of the metal line systems.

\begin{table}[!ht]
  \caption{Summary of the observations.}
  \smallskip
  \begin{center}
    {\small
      \begin{tabular}{clccrc}
	\tableline
	\noalign{\smallskip}
	spectrograph & date & $t_{\mathrm{exp}}$(ks) & resolution & $S/N$ & wavelength (\AA) \\
	\noalign{\smallskip}
	\tableline
	\noalign{\smallskip}
        FUSE       & Feb/May 2003    & 203  & 15\,000 & 5   & 1000 -- 1180 \\
	Keck/HIRES & $^{\mathrm{a}}$ & 84.2 & 37\,500 & 100 & 3670 -- 5880 \\
	HST/STIS   & July 1998       & 76.5 & 45\,800 & 3   & 1230 -- 1500 \\
      \end{tabular}
    }
  \end{center}
  \begin{list}{}{}
    \item[$^{\mathrm{a}}$] datasets published in \citet{songaila98} and \citet{simcoeetal02}
  \end{list}
  \label{observations}
\end{table}

To estimate the continuum in the FUSE portion we have taken low resolution HST/STIS data in May 2003. 
The covered wavelength range (1150 -- 3200\,\AA) includes an overlap with the FUSE data, where the two spectra match despite the UV variability of the QSO.
We extrapolate the continuum from the STIS data correcting for extinction, the seven LLS, and considering the intrinsic QSO power law continuum.

\section{Modelling the metal line systems}

\begin{figure}[!ht]
  \plotfiddle{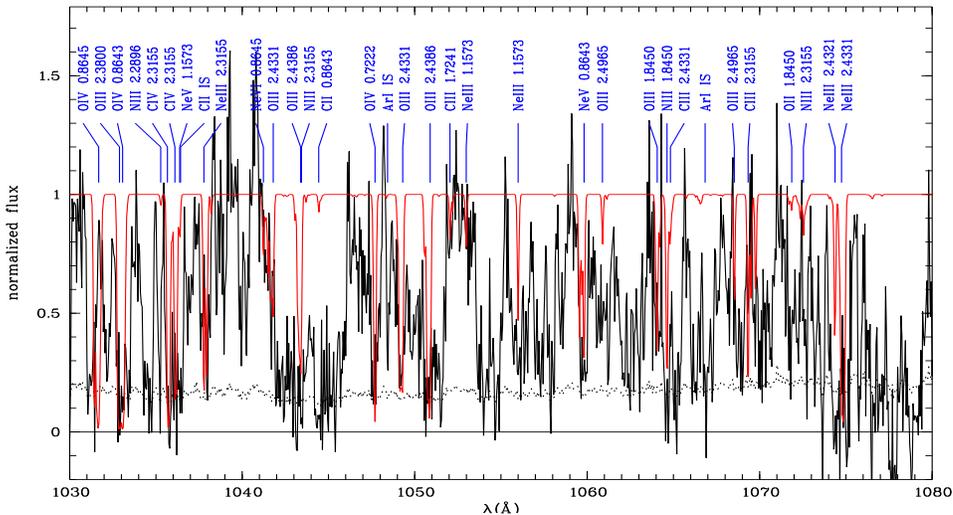}{6.44cm}{-90}{47}{37}{-200}{215}
  \caption{Portion of the predicted metal lines in the FUSE part of the spectrum in comparison to the data.}
  \label{fig1}
\end{figure}

We derive models for 18 metal line systems with redshifts in the range of $0.2140 \le z \le 2.5786$ using the photoionization code CLOUDY \citep{cloudy}.
A Haardt--Madau UV background at the appropriate redshifts \citep{HM96} serves as ionizing radiation field in most of the cases. Three systems require a power law continuum.
The ionization parameter is fixed on the basis of column density ratios of two ionization stages of at least two different elements.
Metallicities and relative abundances are scaled to match the observed column densities.
Using this procedure we describe the observed column densities and get a prediction for the distribution of metal lines in the FUSE wavelength range.
As can be seen from Fig. \ref{fig1} some rather strong features can be identified with metal lines.

\section{Results}

\begin{figure}[!ht]
  \plotfiddle{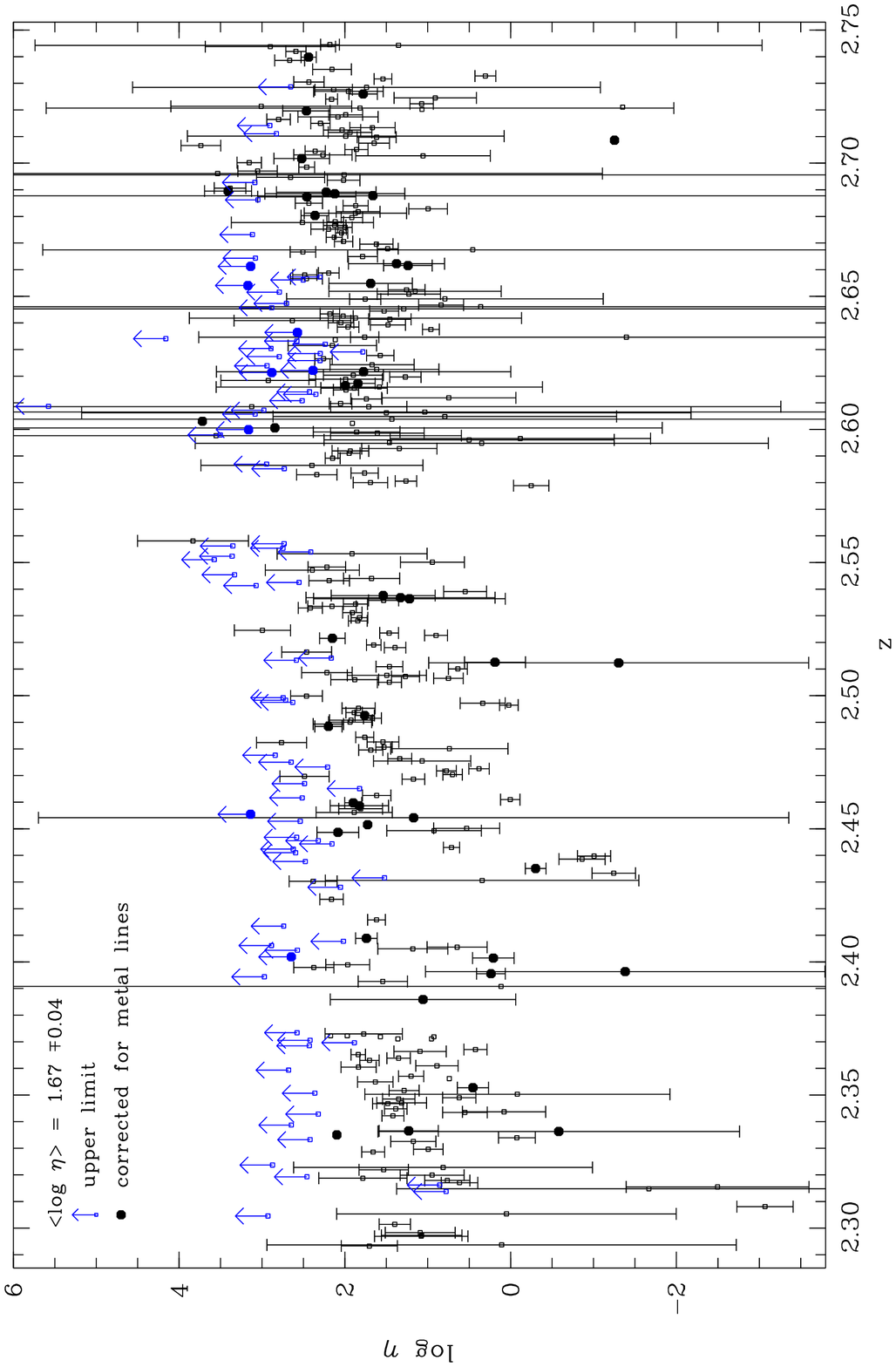}{2.45cm}{-90}{32}{33}{-200}{100}
  \plotfiddle{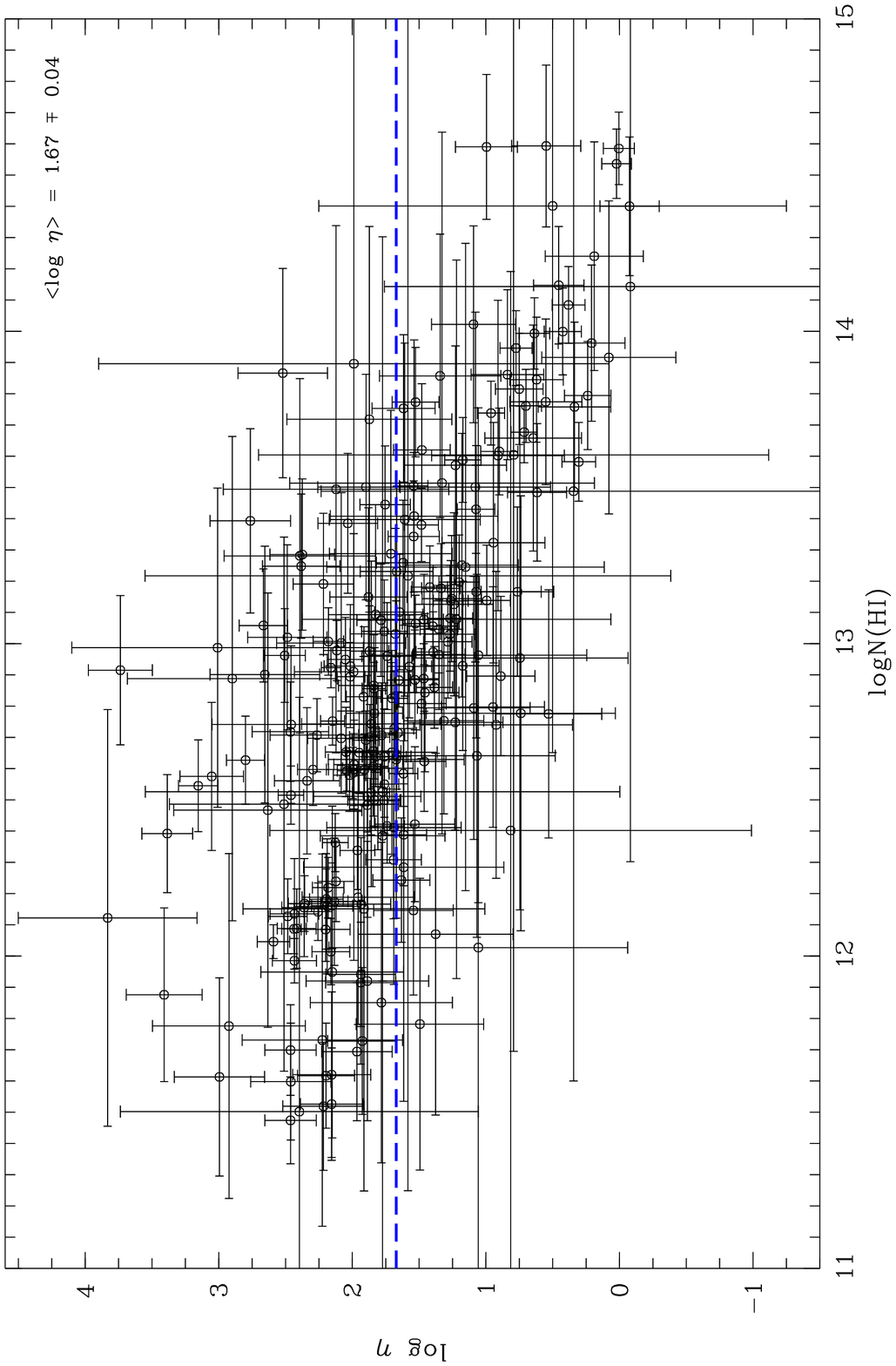}{2.45cm}{-90}{20}{33}{45}{183}
  \caption{Left panel: distribution of $\log\eta$ with redshift considering metal line absorption. Values that have been biased by unrecognized metal line absorption are marked as filled circles. Right panel: Correlation of $\log\eta$ with \ion{H}{1} column density. The dashed line indicates the mean value.}
  \label{fig2}
\end{figure}

We fit all \ion{H}{1} absorbers with Doppler profiles and model the corresponding \ion{He}{2} features at the same redshifts assuming pure turbulent broadening. If necessary, we add \ion{He}{2} lines without \ion{H}{1} counterpart.
The result for the column density ratio $\eta$, taking into account the metal line prediction, is shown in the left panel of Fig. \ref{fig2}. 
The distribution resembles the result of HE~2347-4342.
The consideration metal absorption lines has no dramatic influence on the result in comparison to a simple model, even though 13.5\,\% of the \ion{He}{2} features would have been biased by metal lines.
The average value $\langle\eta\rangle$ decreases by $\sim$ 17\,\% and we have to add 7\,\% less \ion{He}{2} lines without \ion{H}{1} counterpart.
The average column density ratio $\langle\log\eta\rangle = 1.67 \pm 0.04$ corresponds to a Haardt--Madau ionizing background, but individual values are in a range from $\sim 1$ to more than 1000, which implies highly fluctuating ionization conditions.
Furthermore, we confirm the correlation between \ion{H}{1} column density and $\log\eta$ (Fig. \ref{fig2}, right panel) as found in previous studies \citep{shulletal04}, where weak \ion{H}{1} absorbers have high $\eta$-values.



\begin{thebibliography}{}

\bibitem[Ferland(1997)]{cloudy}
  Ferland, G. 1997, A Brief introduction to Cloudy (Internal Rep., Lexington: Univ. Kentucky)
\bibitem[Haardt \& Madau(1996)]{HM96}
  Haardt, F. \& Madau, P. 1996, ApJ, 461, 20
\bibitem[Kriss et al.(2001)]{krissetal01}
  Kriss, G. A., Shull, J. M., Oegerle, W., Zheng, W., Davidsen, A. F. et al. 2001, Science, 293, 1112
\bibitem[Simcoe et al.(2002)]{simcoeetal02}
  Simcoe, R., Sargent, W., \& Rauch, M. 2002, ApJ, 578, 737
\bibitem[Songaila(1998)]{songaila98}
  Songaila, A. 1998, AJ, 115, 2184
\bibitem[Shull et al.(2004)]{shulletal04}
  Shull, J. M., Tumlinson, J., Giroux, M. L., Kriss G. A., \& Reimers, D. 2004, ApJ, 600, 570

\end{thebibliography}
\end{document}